\documentclass[reqno,12pt]{article}
\usepackage{amsmath,amsfonts,amssymb,amsthm,amstext,amscd,eucal,srcltx}
\usepackage[all]{xy}
\usepackage{epsfig,graphicx,bm}
\usepackage{epstopdf, epsf}
\usepackage{dcolumn}
\usepackage{enumerate}
\makeatletter \@addtoreset{equation}{section}

\makeatletter\renewcommand\section{\@startsection {section}{1}{\z@}%
                                   {-3.5ex \@plus -1ex \@minus -.2ex}
                                   {2.3ex \@plus.2ex}%
                                   {\normalfont\large\bfseries}}
\renewcommand\subsection{\@startsection{subsection}{2}{\z@}%
                                     {-3.25ex\@plus -1ex \@minus -.2ex}%
                                     {1.5ex \@plus .2ex}%
                                     {\normalfont\bfseries}}

 \newcommand{\be}{\begin{equation}}
 \newcommand{\ee}{\end{equation}}
 \newcommand{\bea}{\begin{eqnarray}}
 \newcommand{\eea}{\end{eqnarray}}

\def\oads{O-AdS$_3$ }
\def\ads{AdS$_3$ }
\def\cft2{CFT$_2$ }

\def\tr{\tilde{r}}

\def\trho{\tilde{\rho}}
\def\tPhi{\tilde{\Phi}}
\def\ttau{\tilde\tau}
\def\tphi{\tilde{\phi}}
\def\tpsi{\tilde{\psi}}
\def\z2{$\mathbb{Z}_2$ }

\newcommand{\bse}{\begin{subequations}}
\newcommand{\ese}{\end{subequations}}

\begin{document}

\begin{titlepage}
\thispagestyle{empty}
\begin{flushright}
{
IPM/P-2010/029  \\
}\end{flushright} \vspace{10 mm}
\begin{center}
    \font\titlerm=cmr10 scaled\magstep4
    \font\titlei=cmmi10 scaled\magstep4
    \font\titleis=cmmi7 scaled\magstep4
     \centerline{\titlerm Orientifolded Locally AdS$_3$ Geometries }

 \vspace{1.0cm}
    \noindent{\textbf{\large{F. Loran$^1$ and M. M. Sheikh-Jabbari$^2$}}}\\
    \vspace{0.8cm}

   $^1${\small \it Department of  Physics, Isfahan University of Technology,
  Isfahan 84156-83111,  Iran}\\
\vskip 2mm
  $^2${\small\it  School of Physics,
 Institute for research in fundamental sciences (IPM),\\ P.O.Box
 19395-5531, Tehran, Iran}\\
\vskip 2mm
 E-mail:  loran@cc.iut.ac.ir,\ \  jabbari@theory.ipm.ac.ir
  \end{center}

  \vskip 2em

  \begin{abstract}
Continuing the analysis of [arXiv:1003.4089[hep-th]], we classify
all locally \ads stationary axi-symmetric  \emph{unorientable} solutions to  \ads Einstein gravity and show
that they are obtained by applying certain orientifold projection on
AdS$_3$, BTZ or  \ads self-dual orbifold, respectively O-AdS$_3$,
O-BTZ and O-SDO geometries. Depending on the orientifold fixed
surface, the O-surface, which is either a space-like $2D$ plane or cylinder, or a
light-like $2D$ plane or cylinder one can distinguish four distinct
cases. For the space-like orientifold plane or cylinder cases these
geometries solve \ads Einstein  equations and are hence locally \ads everywhere except at the
 O-surface, where there is a delta-function source. For
the light-like cases the geometry is a solution to Einstein equations
even at the O-surface. We discuss the causal structure for static, extremal
and general rotating O-BTZ and O-SDO cases as well as the geodesic
motion on these geometries. We also discuss orientifolding
Poincar\'e patch AdS$_3$ and AdS$_2$ geometries  as a way to geodesic completion of these spaces and comment
on the $2D$ CFT dual to the O-geometries.

  \end{abstract}

\end{titlepage}

\tableofcontents


\section{Introduction }\label{int}

Three dimensional gravity, due to its relative simplicity compared
to higher dimensional gravity theories, has been used as a
laboratory to address questions about quantum gravity. Three
dimensional Einstein gravity on the flat space background has
neither propagating gravitons nor nontrivial (black hole) classical
solutions, its partition function has been computed noting that this
theory is in fact an SL(2, R) Chern-Simons theory \cite{Witten-88}.
Addition of a negative cosmological constant to the theory brings
the possibility of having black hole solutions, the BTZ black holes
\cite{BTZ1, BTZ2}. BTZ black holes has appeared as the prime arena
for addressing black hole thermodynamics puzzles and seeking
statistical mechanical resolutions in terms of the proposed dual
$2D$ CFT.

In three dimensions Riemann tensor is completely specified by the
Ricci tensor \cite{Carlip-Book-review} and as such Einstein
equations imply that all the solutions to the pure \ads Einstein
gravity should be locally AdS$_3$. Therefore, the only option for
solutions other than (global) \ads geometry is to orbifold \ads by a
subgroup of its $O(2,2)$ isometry group. And for the latter, as we
will review briefly in section 2 and discussed in \cite{BTZ1, BTZ2,
self-dual-orbifold}, besides the trivial (space-like) \ads orbifolds
AdS$_3/\mathbb{Z}_k$ there seems to be only two possibilities
leading to either BTZ black hole or the \ads self-dual-orbifold
(SDO). (We are of course excluding the pathologic geometries which
involve closed time-like curves, CTC's, not dressed by an event
horizon.)

In \cite{LS1} we revisited the problem of classification of
solutions to \ads gravity and noted that \ads geometry besides the
orientation preserving $SO(2,2)$ isometry is invariant under an
orientation changing \z2 and one may construct a new class of
solutions by orbifolding (orientifolding)  this $\mathbb{Z}_2$.
In this way we constructed the new class of BTZ geometries,
\emph{orientifold-BTZ} or O-BTZ geometries. There are various
possibilities for the choice of this \z2 but there is only one possibility
which does not change the orientation on the $2D$ causal (conformal) boundary of the
AdS$_3$. This \z2 commutes with the BTZ orbifolding and hence the orientifolding and BTZ orbifolding can be performed at the same time.

In this work we extend the analysis of \cite{LS1} and construct all
the orientifold \ads geometries, \emph{O-geometries} for short, with
this choice of $\mathbb{Z}_2$. As we will show the O-geometries are
necessarily of the form of orientifolded AdS$_3$ (O-AdS$_3$),
orientifolded BTZ (O-BTZ) or orientifolded
self-dual-AdS$_3$-orbifold (O-SDO). These O-geometries, are hence
locally AdS$_3$ by construction. There is, however,  a special
locus, the \emph{O-surface}, the fixed $2D$ surface of the
orientifold operation we perform. The O-surface, being fixed point
of the orientifold projection, should in fact be viewed as the
boundary (of course not a conformal, causal boundary) of the
O-geometries. The O-surface is in fact a Cauchy surface. As we will
explicitly show this orientifold fixed locus is a $2D$ space-like
surface with topology of $R^2$ for the case of \oads while it is a
cylinder for generic O-BTZ. This space-like fixed plane (cylinder)
becomes light-like when we approach the causal boundary of the
geometry. For extremal O-BTZ the O-surface is light-like. For the
O-SDO there are two possibilities of having space-like or light-like
$2D$ fixed cylinders.

As discussed O-geometries are locally \ads everywhere away from the O-surface. One may still study the curvature of the O-geometry at the O-surface which should be viewed as the ``end locus'' of the geometry. This may be carried out if we go to  the ``covering space'' of the projection and extend the space to behind the O-surface. One may then use the Israel matching conditions \cite{Israel} or its refined formulation of \cite{Mansouri-Khorrami} (which is reviewed in Appendix C) to compute the Ricci at the O-surface. The Ricci, which as we will show has a delta-function jump at the O-surface, may be associated to a stress tensor of an ``orientifold plane'' (cylinder) sitting at the O-surface. We note, however, that for the light-like O-surface the
Ricci is continuous.

In this paper we study in some detail the orientifolded locally \ads
geometries as classical (Einstein) gravity backgrounds. In section
\ref{revisit} we review the known and well-established solutions to
pure \ads Einstein gravity. We review construction of BTZ and
self-dual \ads orbifold solutions. In section 3 we argue for the
possibility of constructing locally \ads \emph{unoriented}
geometries (O-geometries), classify all of them and analyze their
causal structure. In section 4 we study geodesic motion on the
O-geometries.  In the last section, the discussion
section, we discuss the relevance of the O-BTZ geometries to the
possible dual $2D$ CFT description for the \ads Einstein quantum
gravity and outline future studies in this direction. In a couple of
appendices we have gathered a summary of useful notations, some
technical details of the computations and some
new  solutions to \ads Einstein gravity coupled to conformal
matter fields.


\section{Orientable solutions to \ads Einstein gravity, a quick
review} \label{revisit}

As mentioned in the introduction all the solutions of \ads Einstein
gravity are locally \ads and may be obtained by modding out the
space by a subgroup of its isometries. In this section we review the
well-known solutions, namely BTZ black holes and the self-dual \ads
orbifold (SDO) which are obtained by modding out the \ads space by a
part of its orientation preserving $SO(2,2)$ isometry group.

\subsection{The BTZ black holes}

A generic rotating BTZ black hole  can be constructed by orbifolding
original AdS$_3$ by the boosts of its  $SO(2,2)$ isometry. In terms
of the embedding space coordinates \eqref{AdS-embedding} that is,%
\be\label{BTZ-identifications}%
\begin{split}
&T_1\pm X_1\equiv e^{\pm \frac{2\pi r_+}{\ell}}(T_1\pm X_1)\ , \cr
&T_2\pm X_2\equiv e^{\pm \frac{2\pi r_-}{\ell}}(T_2\pm X_2)\ .
\end{split}
\ee%
where  $r_+>r_-\geq 0$. $r_+=r_-$ case, corresponding to the extremal
(or massless for $r_+=r_-=0$) BTZ black hole is in a different class
and cannot be constructed through \eqref{BTZ-identifications}. For
$r_-=0$, the static BTZ black hole, the above orbifolding has a
fixed line at $T_1=X_1=0,\ T_2^2-X_2^2=\ell^2$ while for generic
$r_-\neq 0$ case the orbifolding is freely acting on \ads and we
have a smooth geometry. In the coordinate system
\eqref{line-element} the BTZ identification
\eqref{BTZ-identifications} is written as%
 \be
 \label{t-phi-identification}%
 (\ttau,\tr,\tphi)\sim (\ttau-2\pi r_-/\ell,\ \tr,\ \tphi+2\pi r_+/\ell).%
 \ee%
The BTZ geometry has two horizons which in our coordinate system
\eqref{line-element} are at $\tr=\ell$ and $\tr=0$. In the BTZ
coordinates (when $r_+\neq r_-$)
\be\label{BTZ-vs-Steif-coordinates}%
\begin{array}{l}
\ttau=\frac{1}{\ell}(r_+\tau-r_-\phi),\\\\
\tphi=\frac{1}{ \ell}(r_+\phi-r_-\tau),\\\\
\tr^2=\frac{\ell^2}{r_+^2-r_-^2}(r^2-r_-^2),
\end{array}
\ee%
metric takes the form %
\be\label{BTZ-coordinates}%
ds^2=\rho^2d\tau^2+\frac{r^2dr^2}{16G^2J^2-\frac{r^2\rho^2}{\ell^2}}+r^2d\phi^2
-8G\ell J d\tau d\phi,
\ee%
where now the identification is only made along the $\phi$
coordinate $\phi\in [0,2\pi]$ and%
\be%
\rho^2=8GM\ell^2 -r^2. %
\ee%

In this coordinate system the outer and inner horizons are located
at $r=r_+$ and $r=r_-$ respectively. The (ADM) mass $M$ and angular
momentum $J$ are given by%
\be
\label{AdS-MandJ}%
M=\frac{r_+^2+r_-^2}{8\ell^2 G},\quad J=\frac{r_+r_-}{4G\ell}\ .
\ee%

We note that the coordinate transformation
\eqref{BTZ-vs-Steif-coordinates} is singular for the extremal
$r_+=r_-$ case. As discussed in \cite{BTZ2}, however, one may still
use \eqref{BTZ-coordinates} for this case.

\begin{figure}[t]
\includegraphics[scale=1]{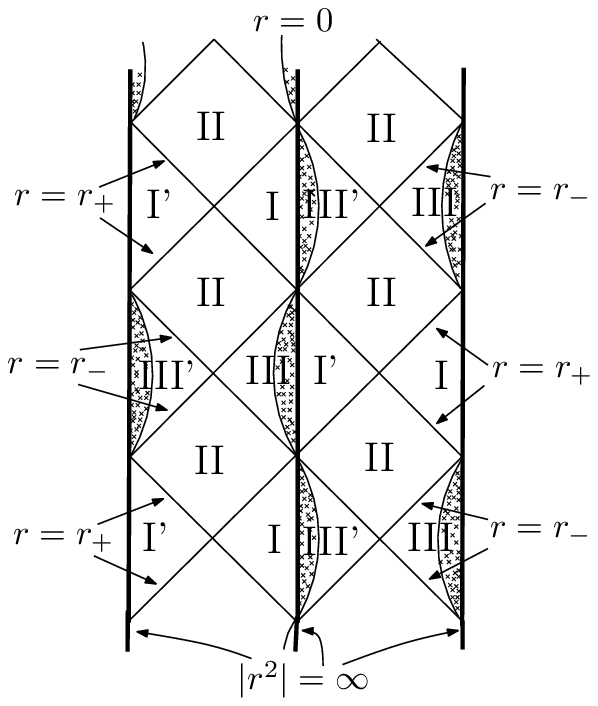}\hspace{20mm}%
\includegraphics[scale=.6]{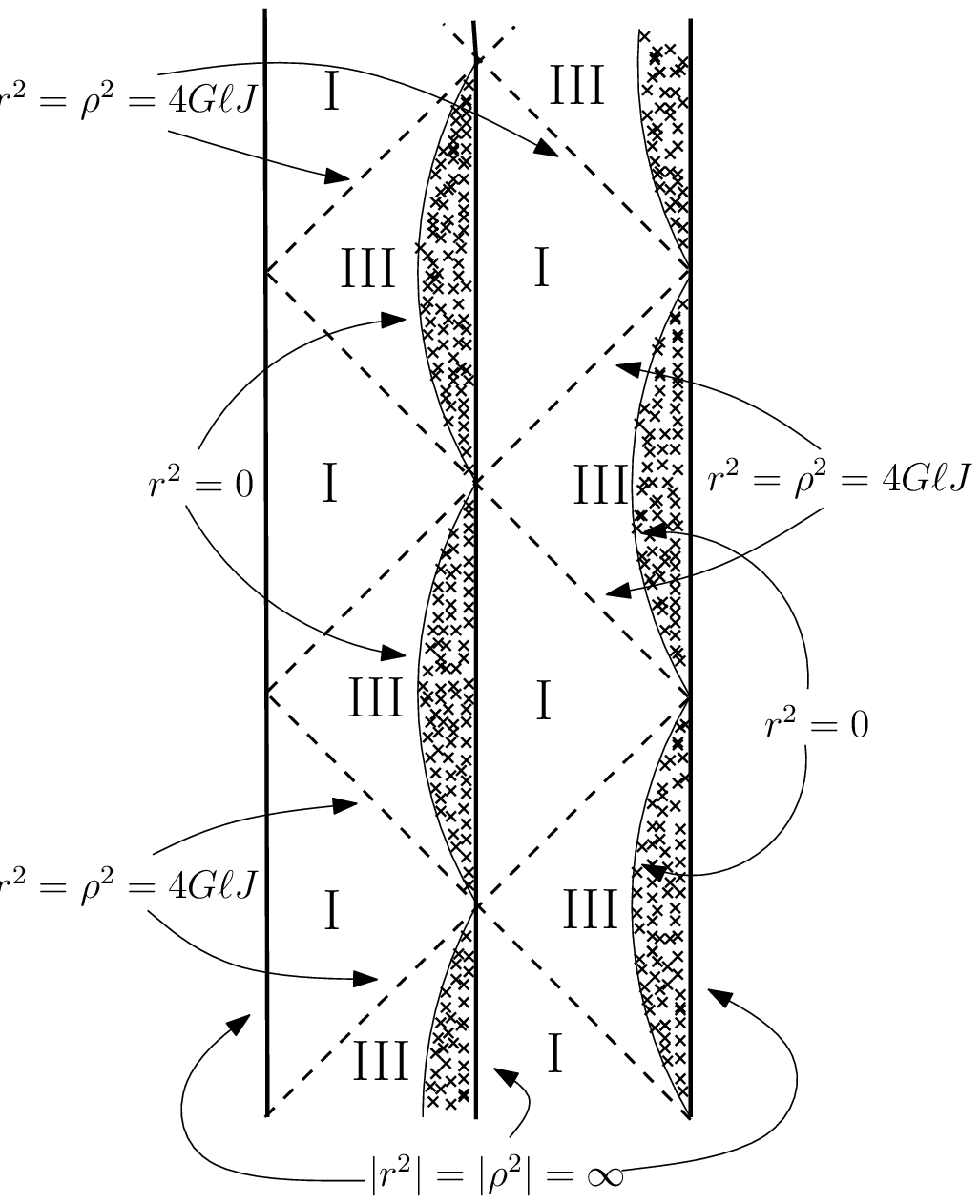}%
\caption{\label{BTZ-Penrose} {\textit{Two Penrose diagrams of BTZ
black holes drawn side-by-side. \textbf{Left figure}: A generic
rotating BTZ.} Regions I, I'  are bounded between the boundary at
$r^2=+\infty$ and the outer horizon at $r=r_+$. Region II is the
region between the two horizons and Regions III, III' are bounded
between the inner horizon (which is also a Cauchy horizon) and boundary at $\rho^2=\infty$. \textit{
\textbf{Right figure}: An extremal BTZ.} Here the region II is
absent and the two horizons coincide. The horizon in this case is
also a Cauchy horizon. In both of the cases Region III may not be
reached by any physical observer from the Region I and  the hatched
area, corresponding to $r^2<0$, which is the region containing closed time-like curves, is cut from the BTZ geometry. As
depicted, this region contains a part of the causal boundary of
original AdS$_3$ \cite{BTZ2}. In the figure  $\phi$ direction has
been suppressed and the $|r^2|=\infty$ lines correspond to
$\phi=0,\pi, 2\pi$. To convey the idea of the suppressed $\phi$ direction and that $r$ coordinate can be
extended past $r^2<0$ we have drawn two Penrose diagrams
side-by-side.}}
\end{figure}

{}As discussed in \cite{BTZ1,BTZ2} the identification
\eqref{t-phi-identification}  leads to closed time-like curves
(CTC's). Recalling metric \eqref{Steif-metric-I}, the CTC's will be generated where%
\be%
{\cal D}=-(\tr^2-\ell^2)r_-^2+\tr^2 r_+^2=\tr^2(r_+^2-r_-^2)+\ell^2
r_-^2
\ee%
which measures the length of the curve closed by the
identifications, becomes negative. With our choice  $r_+> r_-\geq
0$ CTC's develop for%
\be\label{CTC-development}%
\tr^2 \leq \tr^2_{CTC}=-\ell^2\frac{r_-^2}{r_+^2-r_-^2} < 0.%
\ee%
That is, in region III (\emph{cf.} Appendix A) we will have CTC's.
In the BTZ coordinates \eqref{BTZ-vs-Steif-coordinates} this happens
at $r^2<0$ ($\rho^2>8GM\ell^2$). To remove inconsistencies arising
from CTC's, as prescribed in \cite{BTZ2}, the $r^2<0$ region is cut
out from the geometry.\footnote{It was argued in \cite{BTZ2} that
for the static case addition of any matter to the BTZ background
turns $r=0$ to a curvature singularity. This latter was confirmed in
\cite{Steif, ortiz-lifschytz} by showing that expectation value of
the energy momentum tensor corresponding to quantum fluctuations of
a matter field added on the BTZ background blows up at $r=0$, a
result which is also supported by semi-classical AdS/CFT treatments
using (space-like) geodesics \cite{probing-inside, Vijay}. One can
then show explicitly that the back-reaction of this energy momentum
tensor on the geometry  creates a curvature singularity at $r=0$
(see Appendix B). For the $J\neq 0$ rotating backgrounds, however,
the situation is different. The energy momentum corresponding to
fluctuations of any matter field will blow up at the inner horizon
$r=r_-$, rather than $r=0$. This result agrees with semiclassical
AdS/CFT analysis based on (space-like) geodesics \cite{Krishnan}.
However, back-reaction of energy momentum tensor given in
\cite{Steif, ortiz-lifschytz} seems to destroy the asymptotic \ads
geometry, making the perturbative analysis of \cite{Steif,
ortiz-lifschytz}, where the back-reaction effects are not accounted
for, inapplicable.} This would render the BTZ black hole geometry as
geodesically incomplete.
Although region
III is not accessible to any observer from region I, points in
region III and II can be related by geodesics. This renders the BTZ
black holes as geodesically incomplete and may cause problems for a
quantum gravity description of BTZ geometries (e.g. using dual CFT
language). Despite the arguments for the necessity of excising the
inner horizon and the region behind it from the geometry (e.g. see
\cite{Vijay,Krishnan}) it is not clear whether it is possible to
carry this out within a (unitary) dual CFT. Penrose diagram of the
BTZ geometry is depicted in Fig.\ref{BTZ-Penrose}.

\subsection{The \ads self-dual-orbifold, SDO}

 BTZ geometries are stationary, axisymmetric, asymptotically \ads
black hole solutions of pure Einstein \ads gravity. There is another
solution of this theory, which is not a black hole and preserves
SL(2,R)$\times$U(1) of the $SO(2,2)$ isometry and similarly to the
BTZ case can be obtained through  orbifolding \ads by the appropriate
element of the $SO(2,2)$ isometry group. These are the so-called
\ads self-dual orbifold (SDO) geometries constructed in
\cite{self-dual-orbifold}, see also \cite{Assad-Joan-Vijay}.

The metric for the SDO geometry can be given as%
\bse\label{SDO-metric}%
\begin{align}
ds^2&=\frac{\ell^2}{4}(d\tilde t^2+d\tilde
\psi^2)+\left(\frac{\trho^2-\tr^2}{2}\right)d\tilde t
d\tilde\psi-4\ell^2\frac{d\tr^2}{\trho^2}\\
&= {\ell^2}( -r^2 d\tau^2
+\frac{dr^2}{r^2})+\frac{\ell^2}{4}(d\psi+2{r}d\tau)^2%
\end{align}
\ese%
where $\psi$ and $\tilde\psi$ are compact direction,
$\psi\sim\psi+2\pi$ and $\tilde\psi\sim\tilde\psi+2\pi$, $r\geq 0$
and as before $\tr^2+\trho^2=\ell^2$. As is seen from
(\ref{SDO-metric}b) SDO is an AdS$_2\ltimes$S$^1$ geometry, with a
manifest SL(2,R)$\times$U(1) isometry. The causal boundary of SDO
are two disconnected cylinders, the circular section of which is
light-like \cite{Assad-Joan-Vijay}. The SDO in the form (\ref{SDO-metric}a) and
(\ref{SDO-metric}b) can respectively be obtained from the near
horizon limit of near extremal BTZ, and extremal BTZ black holes
\cite{BDSS}.

\section{Unoriented solutions to \ads Einstein
gravity}\label{Unoriented-section}

So far we have reviewed the well-established locally \ads solutions.
These solutions are obtained by orbifolding \ads by a subgroup of
its orientation preserving $SO(2,2)$ isometries. These subgroups,
were also chosen such that they preserve the orientation on the $2D$
boundary of the space which has the topology of $R^{1,1}$ or
$R^1\times S^1$. In this section, following \cite{LS1}, we construct
all unoriented locally \ads geometries. We classify these
geometries, the O-geometries, upon the condition that they have an
orientable $2D$ conformal boundary and show that there is an
O-geometry for any given orientable class of solutions discussed in
previous section.

The orientifold projection we make has a $2D$ fixed surface, the
O-surface. One can then distinguish two classes of O-geometries: those
with a space-like O-surface which will be discussed in sections
\ref{Oads-section}, \ref{OBTZ-section} and \ref{OSDO-section}, and
the other with light-like O-surface to be analyzed in
sections \ref{OBTZ-section}, \ref{Opoincare-section} and \ref{OSDO-section}.

\subsection{Orientifolded AdS$_3$, O-\ads}\label{Oads-section}

To start we  construct  O-AdS$_3$ which is obtained from
orientifolding AdS$_3$ by a specific \z2 part of its isometries.
This \z2 is most simply demonstrated in terms of the coordinate
frame \eqref{Steif-embedding}, acting by exchange of $\rho_1$ and
$\rho_2$, while keeping $\chi_i$. In terms of coordinates employed
in \eqref{Steif-metric-I} or \eqref{Steif-metric-II} the \z2 is
simply changing $\tr$ and $\trho$.  The metric for O-\ads hence
takes the form
\be\label{O-ads3}%
\begin{split}%
ds^2&= (\trho^2\theta(\tPhi)+\tr^2\theta(-\tPhi)) d\tilde t^2-\ell^2
\frac{d\tr^2}{\trho^2}+(\tr^2\theta(\tPhi)+\trho^2\theta(-\tPhi))d\tphi^2\cr
    &=(-|\tPhi|+\frac{\ell^2}{2}) d\tilde t^2
+\ell^2\frac{d\tPhi^2}{4\tPhi^2-\ell^4}+(|\tPhi|+\frac{\ell^2}{2})d\tphi^2
\end{split} %
\ee%
in coordinate system \eqref{line-element}, where
$\tPhi=\tr^2-\ell^2/2$ and $\theta(x)$ is a step-function: it is
zero for $x<0$, is one for $x> 0$ and is 1/2 for $x=0$. We note that \eqref{O-ads3} gives the metric for a double cover of O-AdS$_3$, \emph{i.e.} \eqref{O-ads3} is the metric of O-\ads in the \emph{covering} \ads space.

Since $\tr^2+\trho^2=\ell^2$, this projection has a fixed locus at
$\tr=\trho=\ell/\sqrt2$. This fixed locus, the O-surface, in the
notations of Appendix A,  falls in region II where $\tilde t$ and
$\tphi$ are both space-like and hence O-surface is  a $2D$
space-like $R^2$-plane and is spanned by $\tilde t$ and $\tphi$. Upon the projection the geometry at the two sides of this O-surface are identified, \emph{i.e.} the O-\ads geometry is defined only in $\tPhi\geq 0$ region and considering $\tPhi\in R$ is like going to the covering space of the (orientifold) projection. The orientifold fixed plane is where the volume-form of the \ads
space shifts sign. As is seen from the metric \eqref{O-ads3} the
volume-form for $\tPhi>0$ ($\tr^2>\trho^2$) region is proportional
to $\tr d\tr$, while for $\tPhi<0$ $(\tr^2<\trho^2)$ region to
$\trho d\trho=-\tr d\tr$, explicitly exhibiting the orientation flip
at $\tr^2=\trho^2$ point.

It
is instructive to study  the space-like O-surface from the global
AdS$_3$ viewpoint. Using the coordinate systems introduced in
Appendix
\ref{AdS3-coordinates}, we have%
\be%
\tr^2=\trho^2=\ell^2/2 \quad \Longrightarrow \quad
\cos2\tau=\sin^2\theta\cos 2\psi\ .%
\ee%
 As we see, close to the boundary ($\theta=\pi/2$) this
leads to $\tau=\pm \psi+ n\pi,\ n\in \mathbb{Z}$ which are
light-like directions on the boundary. At the center of \ads
($\theta=0$) this is a surface extended in $\psi$ direction and
sitting at $\tau=n\pi/4,\ n\in\mathbb{Z}$. One can check that this
surface is space-like everywhere in the interior of the \ads covering space while
becomes light-like at the boundary.

The metric is obviously locally \ads at any point away from the
O-surface at $\tPhi=0$ and hence a solution to pure \ads Einstein
gravity. By going to the covering space, one may compute the
curvature at the fixed O-surface.\footnote{This is somewhat similar
to the procedure carried out in \cite{Solodukhin} for computing the
curvature at the tip of an orbifold. This point will be discussed
further in section 5.} Metric, by construction, is continuous at the
O-surface. The Ricci tensor, however, is not continuous and has a
jump. One should then analyze the Israel matching conditions
\cite{Israel}. For the latter we use the formulation developed in
\cite{Mansouri-Khorrami}, which
is reviewed in Appendix \ref{MK-review}, and  arrive at%
\be\label{Ricci-jump}%
\breve{R}_{\mu\nu}=\ell^2\ {\rm \it diag}(1,0,-1)\ \delta(\tPhi)\ ,%
\ee%
in $(\tilde t,\tr,\tpsi)$ frame, for the jump in Ricci
tensor.\footnote{Note that the Ricci scalar is continuous and does
not have a jump.} Therefore, to account for the jump
\eqref{Ricci-jump} at the fixed O-surface one may  introduce a
source $S_{\mu\nu}$ on the right-hand-side of Einstein equations. Noting that
$\breve{R}_{\mu\nu}$ is traceless (recall that at $\tPhi=0$ $\tilde
t\tilde t$ and $\tpsi\tpsi$ components of the metric
are equal) one readily obtains%
\be%
|\alpha| S_{\mu\nu}=\frac{\ell^2}{8\pi G}\ {\rm \it diag}(1,0,-1)\
\delta(\tPhi)\ , \qquad
|\alpha|=\sqrt{|g^{\mu\nu}\partial_\mu\tPhi\partial_\nu\tPhi|}=\ell.%
\ee%
One may associate $S_{\mu\nu}$ to a ``space-like orientifold plane''
with tension $T$ sitting at $\tPhi=0$, i.e.
\be%
S_{\mu\nu}=\sqrt{det g_2}\ T\ diag(1,0,-1)\
\delta(\tPhi)\ , %
\ee%
where $\sqrt{det g_2}=\ell^2/2$ is the determinant of induced metric
on the orientifold plane, and
\be\label{tension}%
T=\frac{1}{4\pi G\ell}.%
\ee%
It is remarkable that the energy momentum tensor of this
orientifold plane (cylinder), $S_{\mu\nu}$, has the following
properties
\be\label{Smunu}%
S^\mu_{\ \mu}=0,\qquad S_{\mu\nu}n^\mu n^\nu=0,%
\ee%
where $n^\mu$ is the time-like vector normal to the worldvolume of
the orientifold plane (cylinder), in our coordinate $\mathbf{n}=\frac{1}{\sqrt{2}}\frac{\partial}{\partial r}$ at the O-surface.

The space-like orientifold plane we have
introduced here, similar to the standard O-planes, is not a
dynamical object and is  located at the orientifold fixed
surface. In our case, unlike the usual O-planes, this fixed surface
is space-like. Despite  this point, we should stress that the conformal
boundary of the O-\ads is still an orientable  Lorentizan $2D$
surface. Moreover, unlike usual O-planes and despite the fact that
we have associated a tension to the O-surface, it does not curve the
space: everywhere away from the O-surface the metric is (locally)
AdS$_3$. The Penrose diagram of O-\ads is the same as the Left
figure in Fig.\ref{O-BTZ-Penrose}, except for the fact that the
suppressed direction $\phi$ is now non-compact.

 It is interesting to note that  the density
of jump of the action in the region II on the either sides of the
O-surface, i.e. action evaluated in  $0\leq \tr^2\leq \ell^2/2 $
region minus its value in $\ell^2/2\leq \tr^2\leq \ell^2$, is equal
to the tension of the
space-like orientifold plane. Explicitly%
\bea\label{Einstein-action-value}%
\Delta S &=&\frac{1}{16\pi
G}\left[\int_{\tr^2=\ell^2/2}^{\tr^2=\ell^2}dV\
(R+\frac{2}{\ell^2})-\int_{\tr^2=0}^{\tr^2=\ell^2/2} dV\
(R+\frac{2}{\ell^2})\right]=\frac{1}{8\pi G\ell} \int d\tilde t
d\tpsi \cr & =& \int T\  \sqrt{g_2}\ d\tilde t d\tpsi .
\eea %
One may also read the above equation in a different way: computing
the value of the gravity action on the \emph{covering} \ads background (using
solution \eqref{Steif-metric-I}) in region II, one would obtain the
same result as $\Delta S$ in \eqref{Einstein-action-value}, which in
turn is equal to the tension of the O-surface.

\subsection{Orientifolded BTZ black holes,
O-BTZ}\label{OBTZ-section}

\begin{figure}[]
\includegraphics[angle=0, width=83.3mm, height=83.3mm]{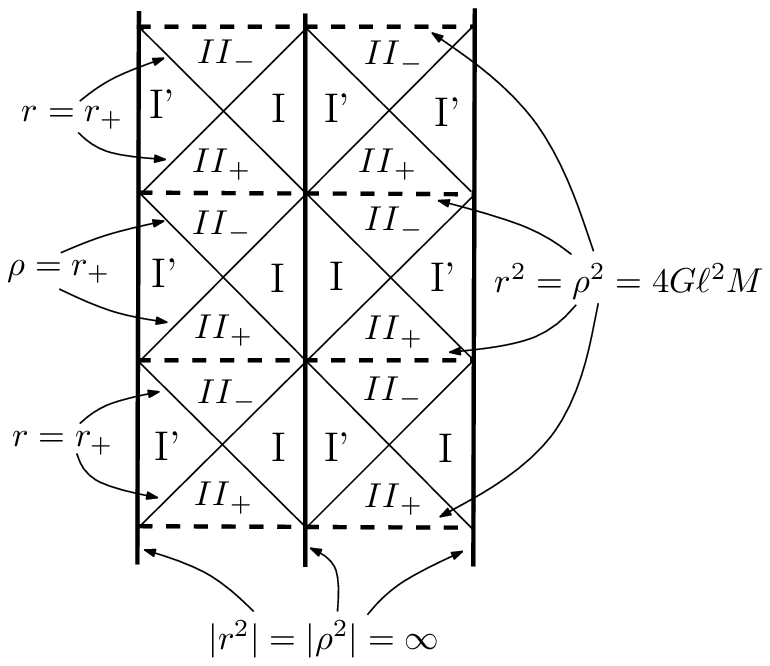}\hspace*{20mm}%
\includegraphics[scale=.67]{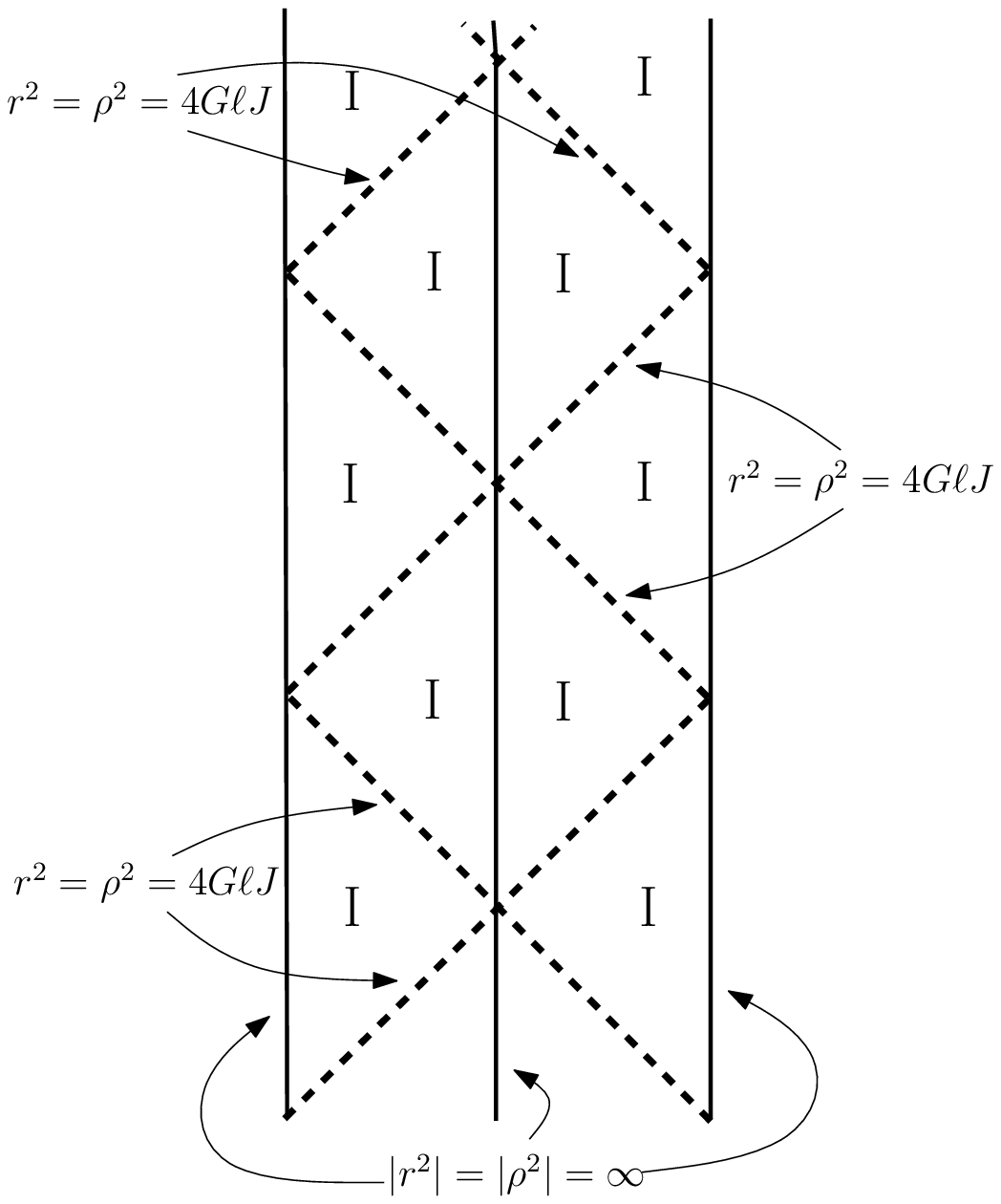}
\caption{\label{O-BTZ-Penrose} {\small\textit{Two Penrose diagrams
of two O-BTZ black holes in the covering space  drawn side-by-side.} \textit{\textbf{Left
figure}: A generic O-BTZ geometry.} The orientifold projection as
expected creates a geometry which is the same on the two sides of
the orientifold fixed surface, the O-surface. The 
O-BTZ is then the square part restricted between the two adjacent
dashed line. In this sense the Penrose diagram is in fact showing
the O-BTZ in the \emph{covering space} of the orientifold projection. In
this case the O-surface is a space-like cylinder and is depicted by
the horizontal thick dashed line. This space-like cylinder, the
circular section of which is suppressed in the Penrose diagram, is
located at $r^2=\rho^2=4G\ell^2 M$ and becomes light-like at the
boundary. The O-BTZ geometry does not have the region III or ``inner
horizon'' region. The causal boundary of O-BTZ geometry is at
$|r^2|=\infty$ which is a cylinder. We point out that  the static
O-BTZ and a generic (non-extremal) O-BTZ geometries have the same
Penrose diagrams. \textit{\textbf{Right figure}: An extremal O-BTZ
geometry.} The orientifold fixed surface is located at the horizon
of the extremal BTZ black hole and, in contrast to the generic O-BTZ
case, is hence a light-like surface; a cylinder the circular
sections  of which is light-like. Again as implied by the
orientifolding, and is explicitly seen in the figure, the extremal O-BTZ geometry is the region I, the triangle bounded between two dashed lines and the vertical $r^2=\infty$ causal boundary, and that the geometry on
the other sides of the dashed line are identical. The diagram is
showing extremal O-BTZ
in its covering space.}}
\end{figure}

Noting that the orientifold projection which led to O-\ads commutes
with the BTZ black hole generating orbifold
\eqref{BTZ-identifications}, one may combine the two and construct
orientifolded BTZ, O-BTZ, geometries. Explicitly, O-BTZ geometry is
obtained by applying the $\mathbb{Z}_2$ projection which in BTZ coordinates
of \eqref{BTZ-coordinates}  takes the form%
\be\label{z2-projection}%
r^2\longleftrightarrow \rho^2,%
\ee%
while keeping $\tau$ and $\phi$, on the BTZ geometry. The double cover of O-BTZ metric (or O-BTZ metric in the covering space)
is then %
\bse\label{OBTZ-metric}
\begin{align}%
\hspace*{-5mm}ds^2 &=[\rho^2 \theta(\Phi)+r^2 \theta(-\Phi)]
d\tau^2-8G\ell J d\tau d\phi +[r^2\theta(\Phi)+
\rho^2\theta(-\Phi)]d\phi^2 +
\frac{r^2dr^2}{16G^2 J^2-\frac{r^2\rho^2}{\ell^2}}\ \\
&=(4G\ell^2M-|\Phi|)d\tau^2-8G\ell J d\tau d\phi
+(4G\ell^2M+|\Phi|)d\phi^2 + \frac{\frac14
d\Phi^2}{\frac{\Phi^2}{\ell^2}-16G^2(\ell^2M^2-J^2)},
\end{align} %
\ese%
where $\theta(X)$ is the step function defined earlier and
\be%
\Phi=r^2-4G\ell^2 M=4G\ell^2 M-\rho^2.%
\ee%
In the coordinate system \eqref{OBTZ-metric}  $\tau$ and $\phi$ are
both dimensionless. It is apparent that metric \eqref{OBTZ-metric}
is invariant under \eqref{z2-projection} which takes $\Phi$ to
$-\Phi$. (It is useful to note that in terms of $\Phi$ the horizons
are sitting at $\Phi=\pm 4G\ell\sqrt{\ell^2M^2-J^2}$.) The volume
element of the
geometry is%
 \be\label{volume-form}\begin{split}%
dV & =\ell d\tau d\phi\ ( \theta(\Phi)\ rdr+\theta(-\Phi)\ \rho
d\rho )\cr &=\ell r d\tau d\phi dr\
(\theta(\Phi)-\theta(-\Phi))=\frac{\ell}{2}d\tau d\phi\ |d\Phi |\ .
\end{split}
\ee%
That is, the two \ads regions on the opposite  sides of the dashed
line in Fig.\ref{O-BTZ-Penrose} have opposite orientations.

O-BTZ geometry is defined in $\Phi>0$ region, where it is locally
AdS$_3$. One would, however, like to study the geometry at the
O-surface $\Phi=0$. With the above choice, metric is clearly
continuous at $\Phi=0$. The
jump of the Ricci tensor is%
\be\label{Ricci-jump-OBTZ}%
\breve{R}_{\mu\nu}=64G^2(\ell^2M^2-J^2)\ {\rm \it diag}(1,0,-1)\ \delta(\Phi)\ ,%
\ee%
in $(\tau,r,\phi)$ frame. This jump is  caused by a space-like
orientifold plane $\Phi=0$ with stress tensor $S_{\mu\nu}=T
\sqrt{\det g_2}\ diag(1,0,-1)\ \delta(\Phi)$, where $\det
g_2=16G^2\ell^2 (\ell^2M^2-J^2)$ is the determinant of the two
dimensional $\tau\phi$ part of metric \eqref{OBTZ-metric} at
 $\Phi=0$, and $T$ is given in \eqref{tension}. It is notable that the tension $T$ is independent of the
the mass $M$ and angular momentum $J$ of the O-BTZ geometry.\footnote{It is instructive to compare our O-BTZ construction and that of (the Lorentzian section of) the geometry constructed in \cite{Kostas}. In the latter, closed time path (CTP) formulation which is often used in real-time non-zero temperature field theory analysis, was applied to the $2d$ CFT's dual to BTZ black holes. The BTZ background was then used as a basis to construct background geometries appropriate for applying gravity dual of CTP formalism. The Lorentizan sector of the geometry discussed in \cite{Kostas} is closely related to our construction in that they cut the BTZ geometry  at a constant $r=r_F$ in the region between the two horizons of BTZ black hole (in their case $r^2_F$ is not necessarily $4GM\ell^2$). In their construction, however, the matching conditions for the geometry is satisfied without any $\delta$-function jump at the junction. This seems to be related to the fact that, due to the CTP formalism, the action on the two sides of the junction should be related by a minus sign. (We note that this is exactly the case for our O-BTZ geometries: the action computed over the $\Phi>0$ and $\Phi<0$ regions of the O-BTZ geometry are equal up to the sign, \emph{cf.}  the last paragraph of section \ref{Oads-section}.) We would like to thank Kostas Skenderis and Balt van Rees for several clarifying email exchanges on this point. }

The O-BTZ black hole, although does not have inner horizon and the
region behind it, has the same line-element as an ordinary BTZ
anywhere away from the O-surface. This in particular implies that
one may associate a Hawking entropy $S_{BH}$ and temperature $T_H$
to the O-BTZ geometry; $S_{BH}$ and $T_H$ have exactly the same
expressions as an ordinary BTZ with the same ADM mass and angular
momentum.

Although the above  analysis works for extremal as well as
non-extremal BTZ cases,  the extremal and massless BTZ cases are
special in some different ways:
\begin{itemize}
\item As depicted in Fig.\ref{O-BTZ-Penrose} for the extremal ($\ell M=|J|$)
case, as well as the massless BTZ case ($\ell M=J=0$), the O-surface
is a light-like cylinder and coincides with the horizon.
\item The jump of curvature $\breve R_{\mu\nu}$
\eqref{Ricci-jump-OBTZ} vanishes for the extremal and massless BTZ
cases. For these cases there is no need to introduce a stress tensor
at the ``light-like orientifold plane''.
\item For the extremal case metric is already invariant under the
\z2 \eqref{z2-projection}, and one only needs to extend the range of
$\Phi$ to negative values. As can be seen from the comparison of the
Right figures in Figs.\ref{BTZ-Penrose}, \ref{O-BTZ-Penrose}, the
only difference between extremal BTZ and extremal O-BTZ is that in
the orientifolded geometry there is no need to cut the shaded region
(associated with the CTC's).
\item  Penrose diagram of massless O-BTZ is the same as a generic
extremal O-BTZ, as depicted in the Right figure in
Fig.\ref{O-BTZ-Penrose}. This is in contrast with the Penrose
diagram of massless BTZ case, which is just a triangle (see figure 5
of \cite{BTZ2}), and is different than that of extremal BTZ. Nonetheless, note that Penrose diagrams of  massive BTZ and massless O-BTZ are the same.
\end{itemize}

Finally we comment that, as in the usual BTZ case, one may still
orbifold the O-BTZ geometry by a $\mathbb{Z}_k$,  corresponding to
reducing the range of $\phi$ coordinate to $\phi\in [0,2\pi/k]$. One
can readily show that the $\mathbb{Z}_k$  orbifolded BTZ black hole
of mass $M$ and angular momentum $J$ is equivalent to a BTZ black
hole (without orbifolding) of mass $M/k^2$ and $J/k^2$. This result
is obviously also true  for O-BTZ.

 \subsection{O-P-AdS$_3$: Orientifolding  \ads on its Poincar\'e
 horizon}\label{Opoincare-section}%
A modified version of the orientifold  projection applied to AdS$_3$
(\emph{cf.} section \ref{Oads-section})  may be applied to \ads in
Poincar\'e coordinates, given by the metric \eqref{Poincare-metric}.
This geometry in a similar way can be extended  to beyond the
Poincar\'e horizon, $u^2>0$ region, by replacing $u^2$ with $|u^2|$.
The metric for the 
O-Poincar\'e-\ads in its covering space is then%
\be\label{O-PAdS3-metric}%
ds^2=\ell^2\left[|u^2|(-dt^2+dx^2)+\frac{du^2}{u^2}\right]=
\left\{\begin{array}{ll}
 \ell^2 u^2(-dt^2+dx^2)+{\ell^2}\frac{du^2}{u^2}&\qquad u^2>0,\\\\
 \ell^2 u^2(dt^2-dx^2)+\ell^2\frac{du^2}{u^2}&\qquad u^2<0,
 \end{array}\right.
\ee%

The O-surface which coincides with Poincar\'e horizon sits at $u=0$ and is a $2D$ light-like $R^2$ plane.
This geometry is similar to that of massless O-BTZ ($\ell M=J=0$)
case, except for the fact that here $x$ direction (corresponding to
$\phi$ direction there) is non-compact. (In the massless O-BTZ the
O-surface is a light-like cylinder.) The Penrose diagram of
O-Poincar\'e \ads is hence the same as the Right figure in
Fig.\ref{O-BTZ-Penrose}.

One may compute the jump of the Ricci tensor for metric
\eqref{O-PAdS3-metric} to be
$$\breve R_{\mu\nu} =4u^4 \ \delta(u^2) \ diag(1,0, -1)=0 ,
$$
in $(t,u,x)$ frame. As a result \eqref{O-PAdS3-metric} is a solution
to Einstein equation.\footnote{Although $\breve R_{\mu\nu}$ vanishes
for this case, one may still formally find the stress tensor
$S_{\mu\nu}$ which creates this jump. Using formalism of
\cite{Mansouri-Khorrami} one finds that $S_{\mu\nu}=\frac{\ell}{4\pi
G} |u^2|\ \delta(u^2)\ diag(1,0,-1)$, which is of course zero.
Nonetheless, noting that square-root of determinant of $t,x$ part of
metric is $\ell^2|u^2|$, the tension for the light-like orientifold
plane is $T=1/(4\pi G\ell)$, which is the same as the O-BTZ cases.}
This is again similar to the massless O-BTZ case.


\subsection{Orientifolded self dual \ads orbifold, O-SDO
 geometry}\label{OSDO-section}
In this section we construct the O-SDO geometry. There are two such
possibilities: to have a space-like O-surface or to have light-like
O-surface. For the former, it is convenient to start with the SDO
metric given in (\ref{SDO-metric}a)
and apply the \z2 orientifold projection%
\be%
\tr\longleftrightarrow\trho,\hspace{1cm}\tilde t\to\tilde t,
\hspace{1cm}\tpsi\to\tpsi.
\ee%
The ``space-like O-SDO'' metric in its covering space is then%
\be\label{O-SDO-metric}%
ds^2=\left\{\begin{array}{ll}\frac{\ell^2}{4}(d\tilde t^2+d\tpsi^2)+
(\frac{\trho^2-\tr^2}{2})d\tilde t
 d\tpsi-4\ell^2\frac{d\tr^2}{\trho^2},&\quad r^2\geq r_*^2\\
\;\;\;\ \\
 \frac{\ell^2}{4}(d\tilde
t^2+d\tpsi^2)+(\frac{\tr^2-\trho^2}{2})d\tilde t
d\tpsi-4\ell^2\frac{d\trho^2}{\tr^2}\ ,&\quad r^2\leq r_*^2
\end{array}\right.
\ee%
where $\tpsi\sim\tpsi+2\pi$ and $r_*^2=\ell^2/2$. The orientifold
fixed surface, which is located at $\tPhi=\tr^2-\ell^2/2=0$, is a
space-like cylinder parameterized by $\tilde t$ and $\tpsi$. The
jump in the
Ricci curvature is%
\be\label{Ricci-jump-SO-SDO}%
\breve R_{\tilde t\tpsi}=\frac{\ell^2}{4}\delta(\tPhi)%
\ee%
and all the other components zero. As we see at $\tr=\trho$ the
off-diagonal part of metric vanishes. As a result $\breve R_{\mu\nu}$
is traceless and  the corresponding energy-momentum tensor
$S_{\mu\nu}$ which creates the Ricci jump \eqref{Ricci-jump-SO-SDO}
satisfies \eqref{Smunu}. As in O-BTZ case, one may associate this
jump to a space-like orientifold cylinder with tension $T=1/(4\pi
G\ell)$.

For the second possibility, the ``light-like O-SDO'', we start with
the metric (\ref{SDO-metric}b) and extend the geometry to negative
$r$ region
by an orientifold projection. This leads to%
\be\label{LO-SOD-metric}%
ds^2={\ell^2}(-r^2 d\tau^2
+\frac{dr^2}{r^2})+\frac{\ell^2}{4}(d\psi+2{|r|}d\tau)^2\ ,\qquad
r\in (-\infty,+\infty).
\ee%
In this case the O-surface is a light-like cylinder and in this
respect is similar to the extremal O-BTZ cases. The jump of the
Ricci at $r=0$, similar to the extremal O-BTZ case, vanishes and
there is no need for the introduction of a new source there. The
conformal boundary of both of the above constructed O-SDO geometries
are two disconnected cylinders with light-like circular sections.

One can show that the metric (for the double cover of) the ``space-like O-SDO''
\eqref{O-SDO-metric} can be obtained from taking the near horizon
limit  over \emph{near extremal} O-BTZ black hole
\eqref{OBTZ-metric} and that ``light-like O-SDO''
\eqref{LO-SOD-metric} can be obtained from the \emph{extremal} O-BTZ
in the near horizon limit. In other words, taking the near horizon
limit commutes with the orientifolding.


 \section{Geodesic motion on O-\ads}

In this section we study the geodesic motion on (the covering space of) the O-geometries and
in particular focus on the behavior of geodesics at the orientifold
fixed surface and establish the geodesic completeness of the
O-geometries which is
also suggested by the Penrose diagrams Fig.\ref{O-BTZ-Penrose}. To
study geodesics on O-\ads it turns out to be more convenient to
choose $\tPhi$ as one of the coordinates, \emph{i.e.}%
\be%
 ds^2=(-|\tPhi|+\frac{\ell^2}{2}) d\tilde t^2
+\ell^2
\frac{d\tPhi^2}{4\tPhi^2-\ell^4}+(|\tPhi|+\frac{\ell^2}{2})d\tphi^2
 \ee
Since the O-\ads
geometry has translation symmetries along $\tilde t$ and $\tphi$,
the geodesics are labeled by the two quantum numbers $E, L$
associated with these symmetries%
\be\label{E-L-tdot-phidot}
\dot{\tilde t}=\frac{d\tilde
t}{ds}=\frac{E}{|\tPhi|-\frac{\ell^2}{2}},\qquad
\dot\tphi=\frac{d\tphi }{ds}=\frac{L}{|\tPhi|+\frac{\ell^2}{2}}\
,\qquad E,L\geq 0,
\ee%
where $s$ is the affine parameter. The geodesic equation then
becomes%
\be\label{geod-OAdS}%
\dot\tPhi^2-\frac{4k}{\ell^{2}}\tPhi^2-4\frac{|\tPhi|}{\ell^2}(E^2-L^2)-(2E^2+2L^2-k\ell^{2})=0 %
\ee%
 where $k=0, +1, -1$ respectively for light-like, space-like and
time-like geodesics. In what follows we will study them separately.\footnote{As the general comment we should stress that whenever a geodesic hits the O-surface, and in the notation of the Left figure of Fig.\ref{O-BTZ-Penrose}, moves from $II_-$ region to $II_+$, it does not leave the O-BTZ geometry; as if it has reentered the geometry in $II_+$ region in the bottom of the O-BTZ square. In other words, all the $II_+$ regions of Fig.\ref{O-BTZ-Penrose} (and similarly for $II_-$, $I$ and $I'$ regions) are identified.\label{goedesic-jump-footnote}}
\subsection{Light-like geodesics}
For the $k=0$ case eq.\eqref{geod-OAdS} reduces to%
\be\label{LL-geod-OAdS}%
\dot\tPhi^2-4\frac{|\tPhi|}{\ell^2}(E^2-L^2)-2(E^2+L^2)=0\ . %
\ee%
The most general solution of the above for $E\neq L$ is
\be%
\tPhi=\sigma B[ (s-s_0)^2-A]\ ,
\ee%
where $\sigma$ is the sign of $\tPhi$ and %
\be
B=\frac{1}{\ell^2}(E^2-L^2)\ ,\qquad A=\frac{E^2+L^2}{2B^2}\ .
\ee%

Depending on the sign of $\tPhi_0=\tPhi(s=0)$, $\dot\tPhi_0=\dot\tPhi(s=0)$ and $B$  there are some different cases which we discuss below:

\begin{enumerate}[i)]
\item
 $E^2>L^2$ and  $\tPhi_0\geq 0,\ \dot\tPhi_0\geq 0$:  The
light-ray starts either in region I  or II and moves away from the O-surface,  toward the \ads boundary. In this case the geodesic never crosses $\tPhi=0$ line.

If $\tPhi_0>\ell^2/2$, it reaches there at finite coordinate time  and bounces back. The motion after the bounce is described by the case ii) below. If $\tPhi_0<\ell^2/2$, it starts in region II and will take infinite coordinate time $\tilde{t}$ to pass to region I.

\item
$E^2>L^2$  and $\tPhi_0>0,\ \dot\tPhi_0<0$. The geodesic starts in either region I or II and
moves toward the O-surface and reaches there at $ s_1=s_0-\sqrt{A}$.
For $s>s_1$ one should use the $\sigma=-1$ branch and continues off to the boundary at $\tPhi=-\infty$. For this case
\be%
\tPhi=\left\{\begin{array}{ll}
B[(s-s_1-\sqrt{A})^2-A] & \qquad  s\leq s_1\\
\;\;\;\; & \;\;\;\ \\
-B[(s-s_1+\sqrt{A})^2-A] & \qquad   s\geq s_1 %
\end{array}
\right. \ee%

As we see $\ddot\tPhi$ at $s=s_1$, where the geodesic crossed $\tPhi=0$, changes sign and jumps by $4(E^2-L^2)/\ell^2$. Recall, however, discussions of footnote \ref{goedesic-jump-footnote}.

\item
$E^2>L^2$  and $\tPhi_0<0,\ \dot\tPhi_0<0$. The geodesic always remains on one side of the O-surface. This case is similar to the case i) and basically the same as what one has on the AdS$_3$. This is of course expected as the $\tPhi<0$ and $\tPhi>0$ regions are related by oreintifold projection.

\item
$E^2>L^2$  and $\tPhi_0<0,\ \dot\tPhi_0>0$. This case is similar to case ii). The geodesic starts in region III or II, moves toward $\tPhi=0$ and passes through where it receives a ``kick''  and continues toward boundary in region I. (Recall discussions of footnote \ref{goedesic-jump-footnote}.) Note, however, that it takes infinite coordinate time to reach to $\tPhi=\infty$.

\item
$E^2<L^2$ case. The light ray oscillates back and forth
with amplitude $A|B|$ and  each oscillation happens in period $4\sqrt{A}$. Each time that the ray reaches
$\tPhi=0$ receives a kick.

\item $E^2=L^2$ case. In this case the geodesic equation does not depend on the sign of $\tPhi$ and in general
$\tPhi=2L\sigma' (s-s_0)$, where $\sigma'=\pm 1$ determines the sign of initial velocity.

\end{enumerate}

 \subsection{Time-like geodesics}
In this case the geodesic equation becomes%
\be\label{geod-OAdS-TL}%
\dot\tPhi^2+\frac{4}{\ell^{2}}\tPhi^2-4\frac{|\tPhi|}{\ell^2}(E^2-L^2)-
(2E^2+2L^2+\ell^{2})=0 %
\ee%
Generic solution to this equation is of the form of
$\cos\frac{2}{\ell}s$ or $\sin\frac{2}{\ell}s$, as in the usual \ads
case. Therefore, the massive particles which follow these geodesics
feel an infinite harmonic oscillator barrier at the boundary and
unlike light-like geodesics will not reach there at a finite
coordinate time $\tilde{t}$; the massive particles oscillate on  paths the
amplitude of which depend on their energy.

The general solution to \eqref{geod-OAdS-TL} is%
\be%
\tPhi=
A\cos\frac{2}{\ell}(s+s_0')+\sigma B %
\ee%
where $\sigma=\pm 1$ is the sign of $\tPhi$ and%
\be%
B=\frac{E^2-L^2}{2}\ ,\qquad A^2=(B+\frac{\ell^2}{2})^2+\ell^2L^2\ .
\ee%
As we see $|B|\leq |A|$ and hence it is possible that $\tPhi$ crosses the $\tPhi=0$ line for generic values of $E$ and $L$. For this case one may choose the origin of
$s$ such that $\tPhi(s=0)=0$ and without loss of generality choose
$A\geq 0$,
leading to \footnote{ $\tPhi$ as a function of $s$ is periodic and one may wonder if this may cause a problem with closed
causal curves.  Noting \eqref{E-L-tdot-phidot}, however, one can show that $\tPhi(t)$ is not periodic.}%
\be%
\tPhi=\left\{\begin{array}{ll}
-2A\sin\frac{s}{\ell}\sin\frac{s+s_0}{\ell} & \qquad \tPhi\geq 0\\
\;\;\;\; & \;\;\;\ \\
-2A\sin\frac{s}{\ell}\sin\frac{s_0-s}{\ell} & \qquad \tPhi\leq 0\
\end{array}
\right. \ee%
where $A\sin\frac{s_0}{\ell}=\frac{\ell}{2}\sqrt{2(E^2+L^2)+\ell^2}$ and we have chosen the solution
such that $\dot{\tPhi}$ is continuous at the O-surface. As we see
the second derivative of $\tPhi$, $\ddot{\tPhi}$, at $s=0$, similar
to the light-like cases, has a jump and changes sign, from
$\frac{4B}{\ell^2}=\frac{2(E^2-L^2)}{\ell^2}$ to minus itself which is the same jump that a light-like geodesic experiences while crossing the O-surface which means it reappears in the region $II_+$ on the bottom of the O-BTZ square (\emph{cf}. footnote \ref{goedesic-jump-footnote}.) 

{}From the above one can conclude that, imposing perfectly
reflecting boundary conditions at the conformal boundary of the
O-\ads geometry all the causal curves can be completely determined
by specifying initial conditions on the orientifold surface at
$\tPhi=0$. In other words $\tPhi=0$ is a Cauchy surface.

\subsection{Space-like geodesics}
In this case we should study%
\be\label{geod-OAdS}%
\dot\tPhi^2-\frac{4}{\ell^{2}}\tPhi^2-4\frac{|\tPhi|}{\ell^2}(E^2-L^2)-(2E^2+2L^2-\ell^{2})=0\ . %
\ee%
Depending on the values of $E$ and $L$ one can recognize two class of solutions\\

\begin{enumerate}[I)]
\item $\cosh$-solutions:
\be%
\tPhi=
A\cosh\frac{2}{\ell}(s+s_0')-B\sigma
\ee%
where $\sigma=\pm 1$ is the sign of $\tPhi$ and %
\be%
B=\frac{E^2-L^2}{2}\ ,\qquad A^2=(B-\frac{\ell^2}{2})^2-\ell^2L^2=(B+\frac{\ell^2}{2})^2-\ell^2E^2\ .
\ee%
In order to have $\cosh$-solution  one should then have $E\geq L+\ell$ or $E\leq |L-\ell|$.\\

\item $\sinh$-solutions:
\be%
\tPhi=
A\sinh\frac{2}{\ell}(s+s_0')-B\sigma
\ee%
where $\sigma=\pm 1$ is the sign of $\tPhi$ and %
\be%
B=\frac{E^2-L^2}{2}\ ,\qquad A^2=\ell^2L^2-(B-\frac{\ell^2}{2})^2=\ell^2E^2-(B+\frac{\ell^2}{2})^2 .
\ee%
$\sinh$-solution, therefore, exist if $|L-\ell|\leq E\leq L+\ell$. In the $\sinh$-solutions $\dot\tPhi$ and $A$ always have the same sign.
\end{enumerate}

Note that in either of the above solutions sign of $A$  can be positive or negative.
Depending on the sign of $\tPhi_0=\tPhi(s=0)$ and $B$ one can recognize some different cases. The solution for
$\sigma=+1, -1$ will respectively be called positive and negative branches.

\begin{enumerate}[I)]
\item $\cosh$-solutions:
\begin{enumerate}[i)]
\item $B \geq 0$: In this case necessarily $A\tPhi_0\geq 0$. If $|A|\geq B$ ($ E+L\leq \ell,\ E^2+L^2\leq \ell^2/2$)
the geodesic always remains in the same positive or negative branch
that it started, assuming its minimum (maximum, if $\tPhi_0<0$)
value at $|\tPhi|=|A|-B$. While if $|A|<B$ ($ E^2+L^2\geq\ell^2/2$)
the geodesic moves toward the O-surface at $\tPhi=0$, passes through
it and
continues toward the boundary at $-\infty$ ($+\infty$ if $\tPhi_0<0$). Recall, however, footnote \ref{goedesic-jump-footnote}.
In the latter case, after a shift in the origin of $s$%
\be\label{SL-cosh-through}%
\tPhi=\left\{\begin{array}{ll}
2|A|\sinh\frac{s}{\ell}\sinh\frac{s-s_0}{\ell} & \qquad A\cdot s\leq 0\\
\;\;\;\; & \;\;\;\ \\
 -2|A|\sinh\frac{s}{\ell}\sinh\frac{s+s_0}{\ell} & \qquad A\cdot s\geq 0\
\end{array}
\right. \ee%
where $A\sinh\frac{s_0}{\ell}=\frac{\ell}{2}\sqrt{2(E^2+L^2)-\ell^2}$.

\item
$B \leq 0$, $A\tPhi_0 \geq 0$: The geodesic remains in the same
branch that it started and does not cross the O-surface. In this
case the geodesic does not distinguish \ads from O-AdS$_3$.

\item
$B<0$, $A\tPhi_0\leq 0$: This case  is possible only  if
$|B|\geq|A|$. The geodesic oscillates around $\tPhi=0$ with
frequency $2|s_0|$ where
$A\sinh\frac{s_0}{\ell}=\frac{\ell}{2}\sqrt{2(E^2+L^2)-\ell^2}$ and
the amplitude $|B|-|A|$:
\be%
\tPhi=\left\{\begin{array}{ll}
2|A|\sinh\frac{x}{\ell}\sinh\frac{s_0-x}{\ell} & \qquad 0\leq x\leq s_0\\
\;\;\;\; & \;\;\;\ \\
2|A|\sinh\frac{x}{\ell}\sinh\frac{x+s_0}{\ell} & \qquad -s_0\leq x\leq 0\
\end{array}
\right. \ee%
where $x=s-2ns_0,\ n\in \mathbb{Z}$ and we have chosen the origin of $s$ such that $\tPhi=0$ at $s=0$.

\end{enumerate}

\item $\sinh$-solutions:
\begin{enumerate}[i)]
\item
If $A\tPhi_0>0$,  geodesic always remains in the same branch and does not cross the O-surface before bouncing off the conformal boundary. These geodesics hence cannot distinguish \ads from O-AdS$_3$.
\item
If $A\tPhi_0\leq 0$, the geodesic then crosses $\tPhi=0$ line once before reaching the conformal boundary at $|\tPhi|=\infty$. For this case, after a shift in the origin of $s$, the geodesic may be given by $\tPhi=A \sinh\frac{2s}{\ell}$. This case is similar to the case discussed in \eqref{SL-cosh-through}.
\end{enumerate}
\end{enumerate}
As one can directly see from \eqref{geod-OAdS}, regardless of $k$ and for all three time-like, space-like and light-like geodesics, when a geodesic passes through $\tPhi=0$ surface, $\ddot\tPhi$ changes sign and jumps by
$4(E^2-L^2)/\ell^2$.

Geodesic analysis on the O-BTZ  is quite
similar to what we have presented above (except for the fact that
$L$ is quantized for O-BTZ) and we do not repeat that here. In particular
the orientifold surface which for the generic O-BTZ geometry is a
space-like cylinder, is a Cauchy surface. For the extremal O-BTZ or
massless O-BTZ case, where this cylinder is light-like, it is a
Cauchy horizon. These can be readily seen from the Penrose diagrams
in Fig.\ref{O-BTZ-Penrose}.


\section{Discussion and Summary}

In this work, continuing analysis of \cite{LS1}, we have made a
classification of O-geometries, the geometries obtained by modding
out asymptotic \ads solutions by a certain orientation changing
isometry. As discussed, this is the only remaining possibility which
completes the set of solutions to \ads Einstein gravity. Our
O-geometries are: 1) asymptotically \ads with $R^{1,1}$ or $R^1\times
S^1$ conformal boundary, 2)  axisymmetric and stationary and are
hence specified with at most two quantum numbers, and 3) the
orientation changing projection is performed such that the
orientation at the $2D$ conformal boundary is preserved. This
latter, which is a natural demand if we are interested in having the
possibility of a dual $2D$ CFT description, leaves us with only one
choice for the orientifold projection. In other words, with the
above three conditions we have exhausted all the possibilities for
O-geometries.

The O-geometries can be classified by i) topology of the O-surface,
which is either a $2D$ plane or a cylinder and ii) by the fact that
it is space-like or light-like. There are therefore, four
possibilities which we discussed in detail in section 3. As discussed
the O-surface is a Cauchy surface (or Cauchy horizon for the
light-like case) and it is sitting behind the horizon (at the
horizon for light-like case). Note that as usual we are imposing
perfectly reflecting boundary conditions at the conformal boundary.

By construction and as can be explicitly seen from Penrose diagrams
in Fig.\ref{O-BTZ-Penrose} the geometries on the other sides of the
O-surface (dashed lines in Fig.\ref{O-BTZ-Penrose}) are exactly the
same. This means that in studying  physics on the O-geometries one
can restrict oneself to only the part of the geometry between the
two adjacent O-surfaces (dashed lines). That is, a generic O-BTZ is the square 
bounded by the conformal boundary and the
horizontal dashed lines and extremal O-BTZ is the triangle bounded by the conformal boundary and the 
$45^\circ$ dashed lines.
This is compatible with the
fact that the O-surface is a Cauchy surface for generic O-BTZ and a Cauchy horizon for extremal O-BTZ. In
this viewpoint our space-time is bounded between O-surfaces; this is
like the standard picture in presence of orientifold planes. 
It is also intriguing to note that Penrose diagram of O-BTZ and the ``O-BTZ square'', is the same as the Penrose diagram of a de Sitter space \cite{de-Sitter}. This may help with formulating very much sought for dS/CFT correspondence.

For the \emph{space-like O-surface cases} we associated a stress
tensor to the O-surface proportional to a delta-function at its
location. (As shown for the light-like O-surfaces there is no need
to associate a tension to the O-surface.) We would like to stress
that despite this, our O-geometries anywhere away from the O-surface
remain locally AdS$_3$. This, in particular, implies that presence
of O-surface cannot be found out using local differential geometry
tools, like curvature invariants and as such one may treat our
O-geometries as solutions to pure \ads Einstein gravity. In this
respect the situation is similar to an ordinary orbifold
singularity: The $R^2/\mathbb{Z}_k$ orbifold, despite of having an
orbifold singularity, say at $r=0$, has vanishing curvature and as
long as curvature invariants are concerned it is a flat space and a vacuum
solution of  Einstein gravity. One may, however, associate a
delta-function curvature to the orbifold geometry \cite{Solodukhin},
where $R=-\frac{1}{\pi} (1-1/k)\delta^2(r)$. This could be put in a
more formal wording employing classification of space-time
singularities given in \cite{Ellis-Schmidt}. The singularity of
O-geometries with space-like O-surfaces is a ``quasi-regular''
singularity and not a curvature singularity. However, there is a
novel difference compared to the case of orbifold singularity or the
singularity of Misner space \cite{Ellis-Schmidt}: in our
O-geometries the orientifold fixed surface is sitting behind the
horizon and not reachable (in finite coordinate time) by the
observer living on the conformal boundary.

One may wonder if there is a dual CFT$_2$ description for the
O-geometries. This question may be approached  from different
viewpoints: taking O-\ads as the vacuum state of a possible dual CFT
and study O-BTZ as thermal states in this CFT, or viewing
orientifolding as a unitary operation in the CFT$_2$ dual to an \ads
and realizing O-geometries as states in the CFT dual to the original
\ads background. The above two viewpoints may become equivalent if
one can show that the CFT$_2$ has two, essentially similar,
independent sectors. In either case, the existence of a dual CFT$_2$
may be anticipated as one may repeat the Brown and Henneaux
\cite{Brown-Henneaux}  analysis for the O-\ads geometries, almost
verbatim, and obtain a Virasoro algebra as its asymptotic symmetry
group with the central charge %
\be\label{BH-central-charge}%
c=\frac{3\ell}{2G}, %
\ee%
where $\ell$ is the AdS$_3$ radius and $G$ is the $3d$ Newton
constant.

The notable feature of O-BTZ geometries is that they do not have inner
horizon and the region behind it. This in particular, as is also seen from the Penrose diagram Fig.\ref{O-BTZ-Penrose}, implies that O-BTZ geometries, unlike the BTZ, do not have a region with CTC.\footnote{
We comment that a similar idea namely, cutting  the region with CTC's and gluing another part
to the geometry for removing the CTC problem has been  previously
discussed, e.g. see \cite{CTC-godel}. Our idea of orientifolding, despite the similarity in using the matching
conditions at the O-surface, is different as in our setting the geometry on the other sides of the O-surface are identified and that the ``junction'' surface in our case should be viewed as the end point of our O-geometry space-time.} Moreover, this is interesting recalling the
instabilities associated with the presence of inner horizons: One
may study  quantum field theory on a BTZ background and compute the
vacuum expectation value (VEV) of the energy momentum tensor
corresponding to vacuum fluctuations of this quantum field. In doing
so, one finds that it blows up at the inner horizon or at $r=0$ in
the static BTZ case \cite{Steif,ortiz-lifschytz}. The back-reaction
of this energy momentum tensor  changes the background BTZ
geometry and turns it to a geometry with curvature singularity (see
Appendix B). One may then  adopt the images method discussed
in \cite{Steif,ortiz-lifschytz,Krishnan} for the O-BTZ. Noting that
the BTZ orbifold projection \eqref{BTZ-identifications} commutes
with the projection \eqref{z2-projection}, one may readily obtain
the expression for the VEV  of the energy momentum tensor for the
quantum fluctuations on O-BTZ background, $\langle
T_{\mu\nu}\rangle_{O-BTZ}$. Explicitly, if we denote the VEV for BTZ
background by $\langle
T_{\mu\nu}(r)\rangle$,%
$$
\langle T_{\mu\nu}\rangle_{O-BTZ}=\left\{\begin{array}{ll} \langle
T_{\mu\nu}(r)\rangle\ , &\qquad r^2\geq \rho^2\\ \;\;\ & \;\;\ \\
\langle T_{\mu\nu}(\rho)\rangle\ , &\qquad \rho^2\geq r^2\
\end{array} \right.
$$
and therefore $\langle T_{\mu\nu}\rangle_{O-BTZ}$ remains finite
everywhere.

In the way of better understanding the O-geometries and their
possible dual CFT$_2$ description one may embed them into a
supergravity or string theory setting. The obvious questions is
first whether our orientifold projection and the (worldsheet)
orientifold projection performed in string theory are the same or
not and whether our O-surfaces are directly related to the
orientifold planes of string theory. A suggestive feature in this
regard is that the tension associated with the O-surface
\eqref{tension} is independent of the details of the O-geometry and
is only specified by the Newton constant $G$ and \ads radius $\ell$
and is inversely proportional to their product. After developing the
setting one may ask if, similarly to the Maldacena-Ooguri construction
\cite{MO}, string theory on O-\ads is solvable or not. Among the
other things string theory setting may lead us to dual CFT$_2$
description of the O-geometries.

 \section*{Acknowledgement}

M.M.Sh-J. would like to thank the Abdus Salam ICTP for the
hospitality where a part of this research was carried out. We would like to thank Kostas Skenderis and Balt van Rees for useful email correspondence.

\appendix
\section{\ads in different coordinate systems }\label{AdS3-coordinates}
\ads is the maximally symmetric Lorentzian $3D$ space with negative
constant curvature
and is a hyperboloid,
\be\label{AdS-embedding}%
 -T_1^2+X_1^2-T_2^2+X_2^2=-\ell^2,
 \ee%
embedded in a four dimensional space ${\mathbb R}^{(2,2)}$, with
metric
 \be
 ds^2=-dT_1^2+dX_1^2-dT_2^2+dX_2^2.
 \ee
Using the above definition one may adopt various coordinate systems
for describing the \ads space. Three of such coordinate systems, the
global AdS coordinates, the ``BTZ-type'' coordinates and the
Poincar\'e patch coordinates are the ones we will be using in this
paper. Here we review the three coordinate systems through solving \eqref{AdS-embedding} and discuss their
relation.

\subsection*{Global coordinates}%
Global \ads coordinates is given by%
\be%
\begin{split}
T_1=\frac{\ell}{\cos\theta}\cos\tau\ &,\quad X_1=\ell\tan\theta\cos\psi\\
T_2=\frac{\ell}{\cos\theta}\sin\tau\ &,\quad
X_2=\ell\tan\theta\sin\psi
\end{split}
\ee%
where $\theta\in [0,\pi/2)$ is the radial coordinate and $\tau\in
(-\infty,+\infty)$ is the global time and $\psi\in (-\infty,
+\infty)$ is the space-like direction which is usually suppressed at
the level of the Penrose diagram. The \ads metric in this coordinate
system is%
\be%
ds^2=\frac{\ell^2}{\cos\theta^2}\left(-d\tau^2+d\theta^2+\sin^2\theta
d\psi^2\right)%
\ee%
The causal boundary of \ads is the two dimensional plane spanned by
$\tau,\ \psi$ sitting at $\theta=\pi/2$.

\subsection*{Poincar\'e coordinates}
These coordinates cover half of the global \ads and their embedding
is%
\be%
\begin{split}
X_1&={\ell}{u} x,\qquad T_1={\ell}{u} t\ , \cr %
T_2-X_2&=\ell u,\qquad
T_2+X_2=\frac{\ell}{u}\left[1+u^2(x^2-t^2)\right]
\end{split}
\ee%
where $u\geq 0$ and $x,t\in \mathbb{R}$. Metric in this coordinate
system takes the form%
\be\label{Poincare-metric}%
ds^2=\ell^2\left[u^2({-dt^2+dx^2})+\frac{du^2}{u^2}\right]\ . %
\ee%
The causal boundary is located at $u=\infty$. $u=0$ is a light-like
direction in the global \ads and is the Poincar\'e horizon. The
metric \eqref{Poincare-metric} is the geometry which appears in the
near horizon limit of D1-D5 system \cite{ads/cft}.

This coordinate system only covers half of the global \ads
because $T_2-X_2\geq 0$. To cover the other half one may use a
similar  coordinate system with $u$ replaced by $-u$. These two
patches would then overlap at the Poincar\'e horizon $u=0$.

\subsection*{BTZ-type coordinates}

This is the coordinate system which is appropriate for constructing
(non-extremal) BTZ
black hole and is given by%
\be%
\rho_i^2=-T_i^2+X_i^2,\qquad  i=1, 2, %
\ee%
where $\rho_1^2+\rho_2^2=\ell^2$ which can be solved as
 \be\label{Steif-embedding}
 \begin{array}{llll}
 T_i=\sqrt\rho_i\cosh\chi_i,&X_i=\sqrt\rho_i\sinh\chi_i,&\rho_i>0,&-\infty<\chi_i<\infty\\
 T_i=\sqrt{-\rho_i}\sinh\chi_i,&X_i=\sqrt{-\rho_i}\cosh\chi_i,&\rho_i<0,&-\infty<\chi_i<\infty.
\end{array}
 \ee
Depending on the sign of $\rho_i$, three distinct regions
 in \ads can be recognized: region I, where $\rho_1>\ell^2$ and
 $\rho_2<0$, region II, where $0<\rho_i<\ell^2$, $i=1,2$, and region III,
 where $\rho_2>\ell^2$ while $\rho_1<0$.
 Defining $\tphi\equiv\chi_1$ and $\tilde t\equiv\chi_2$, one verifies that
 in the region I, the Killing vectors $\partial_{\tilde t}$ and $\partial_{\tphi}$ are time-like
 and space-like respectively, while they are space-like and time-like in region
 III. In region II, both of these Killing vectors are space-like.

In regions I and II, defining $\sqrt\rho_1=\tr$, the \ads metric
becomes
 \be\label{Steif-metric-I}
 ds^2=(\ell^2-\tr^2)
 d\tilde t^2+ \ell^2\frac{d\tr^2}{{\tr^2}-{\ell^2}}+\tr^2d\tphi^2,
 \ee
 where $\tilde t=\chi_1, \ \tphi=\chi_2$. For regions II and III
 $\trho=\sqrt\rho_2$ and the metric takes the form
 \be\label{Steif-metric-II}
 ds^2=\trho^2
 d\tilde t^2+\ell^2\frac{d\trho^2}{{\trho^2}-{\ell^2}}+(\ell^2-\trho^2)d\tphi^2.
 \ee
Furthermore one can use the identity
 \be
 \tr^2+\trho^2=\ell^2,
 \ee
 to show that
 \be
 dR^2\equiv\frac{d\trho^2}{{\trho^2}-{\ell^2}}=\frac{d\tr^2}{{\tr^2}-{\ell^2}}.
 \ee
 So, in region II, the metric can be given by the following line
 element,
 \be
 ds^2=\trho^2d\tilde t^2+dR^2+\tr^2d\tphi^2.
 \label{line-element}
 \ee
 If we extend the $\tr$ coordinate to region III, where $\tr^2<0$ and
 similarly extend the $\trho$ coordinate to region I, then the \ads
 metric in regions I and III can be given by the same line element
 as (\ref{line-element}).


\section{On solutions of \ads Einstein gravity with conformal matter}

Here, we solve \ads Einstein equations for the stress tensor which is
relevant for studying back reaction of the vacuum expectation value
of stress tensor corresponding to vacuum fluctuations of a
conformally coupled scalar field theory on static BTZ background,
discussed in \cite{Steif,ortiz-lifschytz}. Let us, however, consider the more general problem of finding
\emph{static} asymptotically \ads geometries coupled to a traceless stress tensor.

Using diffeomorphisms one can always bring any $3D$ static metric to the form
\be%
ds^2=-h(r)d\tau^2+\frac{dr^2}{N(r)}+r^2d\phi^2, %
\ee%
where  $\tau$ assume values in $(-\infty,\infty)$ while $\phi$ can range over $(-\infty,\infty)$ or can be periodic $\phi\in [0,2\pi]$ (for the BTZ case). We
would like to solve the Einstein field equations
$$
R_{\mu\nu}-\frac{1}{2} R
g_{\mu\nu}=\frac{1}{\ell^2}g_{\mu\nu}+8\pi G T_{\mu\nu},%
$$
in the presence of a
traceless energy-momentum tensor
\be%
T^\mu_{\ \nu}=\frac{1}{8\pi G}\ diag(T_1,T_2,-T_1-T_2), %
\ee%
where $T_1,T_2$ are only functions of $r$, subject to a boundary condition, %
\be\label{asymp-values}%
 h(r),\ N(r)\to (r^2/\ell^2)\hspace{1cm}{\rm as}\hspace{1cm}r\to\infty.
\ee%

The three independent field equations are%
\begin{equation}
\begin{split}
&\frac{h'}{ h}(T_2-T_1)+2(T'_2+\frac{2T_2}{ r}+\frac{T_1}{ r})=0,\\
&N'=2r(\frac{1}{\ell^2}+T_1),\\
&N\frac{h'}{ h}=2r(\frac{1}{\ell^2}+T_2).
\end{split}
\end{equation}
To solve the above equations for the four unknowns, $h,\ N,\ T_1,\
T_2$ we need to assume a relation between the variables. We will consider three interesting cases below.

\begin{itemize}
\item $T_2=0$: One can readily see that equations yield $T_1=0$ and $N=h=\frac{r^2}{\ell^2}$.

\item $T_1=0$: In this case $N=\frac{r^2}{\ell^2}-M$. There are two cases, either $T_2=0$ which basically reduces to the previous case or $T_2\neq 0$ for which
$T_2^2 r^4 h=1$, where $M$ is an integration constant.
Replacing $h=r^2f(r)^2$, then%
\be\label{f-T1=0}%
(\frac{r^2}{\ell^2}-M)f'=\frac{M f}{r}+\frac{1}{r^2}%
\ee%
and $f(r)=1+\sum_{n=1}\frac{a_n}{r^n}$. Plugging the $f$ expansion into \eqref{f-T1=0} one can compute $a_n$ as a function of $M$. For the first three coefficients one finds $a_1=0$, $a_2=-\frac{\ell^2 M}{2}$ and $a_3=-\frac{\ell^2}{3}$. To this order $T_2=\frac{1}{r^3}-\frac{a_2}{r^5}-\frac{a_3}{r^6}+\cdots$. One can also find a closed form for the solution of  \eqref{f-T1=0}. As it is not illuminating we will not present it here. This solution has curvature singularity at $r=0$.

\item $T_1=T_2$ which is relevant for the quantum fluctuations
mentioned above \cite{Steif}. The solution for this case is, %
\be T_1=T_2=\frac{A}{ r^3}, %
\ee%
and
\be%
 N=f=\frac{r^2}{ \ell^2}+\frac{2A}{ r}-M,
\label{backraction}
 \ee
where $A, M$ are  arbitrary constants. As a result of the back
reaction,  $r=0$ in the original static BTZ geometry (which was
the orbifold singularity of the BTZ construction) now turns to a
curvature singularity.

\end{itemize}

\section{Review of matching conditions in Einstein gravity}\label{MK-review}

For completeness  we present the refinement of Israel matching
conditions \cite{Israel} developed by Khorrami and Mansouri
\cite{Mansouri-Khorrami}. Let us suppose that $\Psi=0$ defines a closed
surface in space-time and hence divides the space-time into the inside
and outside regions, specified by $\Psi<0$ and $\Psi>0$ respectively. Let us
denote metric on the other sides of this matching surface by
$g_{\mu\nu}^{>}$ and $g_{\mu\nu}^{<}$ and suppose that they are
solving Einstein equations in those regions. The metric for the
whole space-time is then given by%
\be\label{metric-glued}%
g_{\mu\nu}=g_{\mu\nu}^{>}\ \theta(\Psi)+g_{\mu\nu}^{<}\ \theta{(-\Psi)}%
\ee%
Moreover, one can always choose
the coordinate system at the junction at $\Psi=0$ such that %
\be%
g_{\mu\nu}^{>}=g_{\mu\nu}^{<}|_{\Psi=0}%
\ee%

The question we are now asking is, under which conditions the metric
\eqref{metric-glued} is a solution to Einstein equations everywhere.
In order to answer this question we need to write down Einstein
equations at the junction and for the latter one needs to compute
jump in the curvature on the other sides of the junction. One can
show that \cite{Mansouri-Khorrami}%
\be\label{KM-jump}%
 \breve{R}_{\mu\nu}=\left(\frac{1}{ 2 g}[\partial_\mu
g]\partial_\nu\Psi-[\Gamma^\rho_{\mu\nu}]\partial_\rho\Psi\right)\delta(\Psi),
\ee%
where $[\Gamma^\rho_{\mu\nu}]$ denotes the jump in the Levi-Civita
connections, and $g$ is the determinant of the metric. This jump in
the curvature should be caused by the stress tensor associated with the
junction through Einstein equations at $\Psi=0$: %
\be%
\breve R_{\mu\nu}-\frac{1}{2}\breve R g_{\mu\nu}=8\pi G\breve
T_{\mu\nu},%
\ee%
where $\breve T_{\mu\nu}=\alpha S_{\mu\nu}\delta(\Psi)$ and%
\be%
\alpha=\sqrt{\left|g^{\mu\nu}\partial_\mu\Psi\partial_\nu\Psi\right|}.
\label{alpha}%
\ee%

\section{O-AdS$_2$}

It is possible to orientifold AdS$_2$ too. One may examine
orientifolding AdS$_2$ in two different coordinate systems with two
different junction conditions. It is straightforward to observe that
it is not possible to  orientifold AdS$_2$ on its Poincar\'e
horizon. This may be seen recalling that the O-AdS$_2$ in Poincar\'e
patch should have a metric like \eqref{O-PAdS3-metric} with the
$dx^2$ term dropped. This will make determinant of metric to be
discontinuous (to be $+1$ in one side and $-1$ on the other)%
. We then
remain with the second option, which we will discuss below.

The metric is%
\be%
ds^2=\frac{1}{\ell^2}\left(\rho^2\Theta(\Phi)+r^2\Theta(-\Phi)\right)
dt^2-\ell^2
\frac{r^2dr^2}{r^2\rho^2}=\frac{1}{\ell^2}\left(-|\Phi|+\frac{\ell^2}{2}\right)
dt^2+\frac{\ell^2}{4}
\frac{d\Phi^2}{\Phi^2-\frac{\ell^4}{4}} . %
\ee%
where $\rho^2=\ell^2-r^2$ and $\Phi=r^2-\ell^2/2$ and $\Theta(x)$ is
the step function: it is $+1$ when $x> 0$, 1/2 for $x=0$ and zero
for $x<0$. The O-surface  is hence at $r^2=\rho^2\equiv
r^2_*=\ell^2/2$.

We should next examine the Israel matching conditions. Using the
formulation developed in \cite{Mansouri-Khorrami} one can compute
the jump of the Ricci tensor:
\be%
\breve R_{\mu\nu}=diag(1, -4) \ \delta(\Phi)\ .%
\ee%

Plugging the above into matching condition
$$
\breve R_{\mu\nu}=8\pi G \breve T_{\mu\nu}= 8\pi G \alpha \
S_{\mu\nu}\delta(\Phi),$$ where
$\alpha=\sqrt{|g^{\mu\nu}\partial_\mu\Phi\partial_\nu\Phi|}=\ell$,
we get
\be%
S_{\mu\nu}=\frac{1}{4\pi G\ell}\ diag(1,-4)\ .%
\ee%
In this case, unlike the \ads cases, $S_{\mu\nu}$ is not traceless. More importantly, we note that this $S_{\mu\nu}$  cannot  be associated with a 0+1 dimensional (orientifold) object at constant
$r$ (note that $S_{rr}\neq 0$). In this sense the O-\ads and O-AdS$_2$ are essentially different.



\end{document}